\documentclass{article}

\usepackage[final]{neurips_2022_ml4ps}

\usepackage[utf8]{inputenc} 
\usepackage[T1]{fontenc}    
\usepackage{hyperref}       
\usepackage{url}            
\usepackage{booktabs}       
\usepackage{amsfonts}       
\usepackage{nicefrac}       
\usepackage{microtype}      
\usepackage{xcolor}         
\usepackage{graphicx}

\title{Can denoising diffusion probabilistic models generate realistic astrophysical fields?}

\author{%
  Nayantara Mudur \\
  Department of Physics \\
  Harvard University\\
  Cambridge, MA, 02138 \\
  \texttt{nmudur@g.harvard.edu} \\
   \And
   Douglas P. Finkbeiner \\
  Departments of Astronomy and of Physics \\
  Harvard University\\
  Cambridge, MA, 02138 \\
\texttt{dfinkbeiner@cfa.harvard.edu} \\
}

\begin{document}

\maketitle

\newcommand\change[1]{{\color{blue}#1}}
\newcommand\verbose[1]{{\color{orange}#1}}

\begin{abstract}
  Score-based generative models have emerged as alternatives to generative adversarial networks (GANs) and normalizing flows for tasks involving learning and sampling from complex image distributions. In this work we investigate the ability of these models to generate fields in two astrophysical contexts: dark matter mass density fields from cosmological simulations and images of interstellar dust. We examine the fidelity of the sampled cosmological fields relative to the true fields using three different metrics, and identify potential issues to address. We demonstrate a proof-of-concept application of the model trained on dust in denoising dust images. To our knowledge, this is the first application of this class of models to the interstellar medium.

\end{abstract}

\section{Introduction}
Generative models of astrophysical and cosmological fields can serve a multitude of purposes. Cosmological simulations take several thousand CPU hours to run and can only be generated for a limited set of parameters. There thus exists a demand for emulators: frameworks that can generate summary statistics \citep{Heitmann:2009cu} or the underlying cosmological fields \citep{PhysRevD.102.103504, jamieson2022field} conditional on an input cosmological / astrophysical parameter set. This can accelerate parameter inference approaches that require evaluating likelihoods at intermediate values. In the context of observed astrophysical fields, statistical descriptions capable of capturing the non-Gaussian nature of the interstellar medium would aid component separation problems encountered in searches for the \textit{B}-mode of the Cosmic Microwave Background \citep{remazeilles2018exploring} and statistical regularization for interstellar dust mapping \citep{green20193d, leike2019charting}.

Score-based generative models \citep{song2020score} have witnessed a surge in interest because of findings that show that they surpass GANs in terms of image fidelity \citep{dhariwal2021diffusion} and their ability to produce realistic images conditional on text inputs \citep{saharia2022photorealistic}. These models learn the gradient of the probability density of the data to learn a generative model of the data distribution. A subset of this class includes denoising diffusion probabilistic models (DDPMs) \citep{ho2020denoising}. \cite{smith2022realistic} used DDPMs to generate galaxy images. \cite{remy2020probabilistic, remy2022probabilistic} used the denoising score matching framework to learn the non-Gaussian component of a prior on weak lensing convergence maps from simulations.

In this work we investigate two applications of DDPMs -- one to simulation data products and one to images of interstellar dust. We train models to generate dark matter density fields from a simulation suite on grids of 64x64 and 128x128 pixels. We then compare five summary statistics between samples from the trained models and the real simulation fields. While we currently generate these fields unconditionally, this benchmarking is an important step toward using these models as emulators, and generating fields conditional on an input parameter vector. We then turn our attention to the real sky, and train a model to generate square patches from an interstellar dust map. We use the trained model as a denoising model and examine how well it can reconstruct the underlying image given a noisy input.

\section{DDPM Background}
\newcommand\xt[1]{\mathbf{x}_{#1}}
\newcommand\qt[1]{q(\xt{#1}|\xt{#1 -1})}
\newcommand\pt[1]{p_\theta(\xt{#1-1}|\xt{#1})}
In this section, we briefly review the denoising diffusion probabilistic model formulation in \cite{ho2020denoising}. A DDPM consists of a forward and a reverse diffusion process, over a fixed number of time steps, T, where $\xt{0}$ is a draw from the image distribution and $\xt{T} \sim \mathcal{N}(0, 1)$. The forward diffusion process is defined according to a variance schedule \{$\beta_t$\}. Thus
$\qt{t} = \mathcal{N}(\sqrt{1 - \beta_t}\xt{t-1}, \beta_t \mathbf{I})$ and $q(\xt{t}|\xt{0}) = \mathcal{N}(\sqrt{\bar{\alpha_t}}\xt{0}, (1 - \bar{\alpha_t})\mathbf{I})$, where $\bar{\alpha_t} = \prod_{t'=1}^t 1 - \beta_{t'}$. The neural network $\mathbf{\epsilon}_\theta (\xt{t}, t)$ parameterizes the reverse diffusion process: $\pt{t} = \mathcal{N}(\mu_\theta (\xt{t}, t), \sigma_t^2\mathbf{I})$ where $\mu_\theta (\xt{t}, t) = \frac{1}{\sqrt{1 - \beta_t}} \big(\xt{t} - \frac{\beta_t}{\sqrt{1 - \bar{\alpha_t}}} \mathbf{\epsilon}_\theta (\xt{t}, t)\big)$ and $\sigma_t^2  = \frac{1 - \bar{\alpha}_{t-1}}{1 - \bar{\alpha}_t} \beta_t$. A simplified loss function is minimized where $L_{t-1} = ||\mathbf{\epsilon} - \mathbf{\epsilon}_\theta(\sqrt{\bar{\alpha}_t}\xt{0} + \sqrt{1 - \bar{\alpha}_t}\mathbf{\epsilon}, t))||^2$ with $\epsilon\sim\mathcal{N}(0, \mathbf{I})$. Thus for each batch a set of timesteps t is uniformly sampled from $t \sim U[1...T]$ to minimize $L_{t-1}$.

\section{Generative Models for Dark Matter Density Fields}
\label{cosmo}

\paragraph{Dataset}
The CAMELS Multifield Dataset (CMD) \citep{villaescusa2022camels} from the Cosmology and Astrophysics with MachinE-Learning Simulations (CAMELS) dataset \citep{villaescusa2021camels} was used for this application. The CMD includes an ensemble of thirteen two-dimensional physical fields for 1000 different simulation parameter vectors where each vector consists of 2 cosmological and 4 astrophysical feedback parameters. Every unique parameter vector has 15 samples and the full data set thus consists of 15000 samples of each physical field. We work with the log (base 10) of the cold dark matter mass density field from the IllustrisTNG hydrodynamical simulation at $z=0$, at two grid sizes (64x64 and 128x128) binned down from the original 256x256. Each side of an image corresponds to $25 h^{-1}$ Mpc. For the 64x64 model, we use fields corresponding to the first 60\% of the parameters as our training data. We augment our fields with rotations and flips, to yield 54000 (9000x6) train fields. For the 128x128 model, we use fields corresponding to the first 70\% of the parameters as our training data. We augment our data with rotations, flips and translational shifts (since the data has periodic boundary conditions). We thus have 252000 (10500x24) train fields. We apply a minmax transform that scales the minimum and the maximum pixel intensity of the full training set to [-1, 1].

\paragraph{Training Details}
We train two models, one at each resolution, for 60k iterations (batch updates). In both cases, we use a forward diffusion process parameterized by a linear variance schedule lying in the range [$10^{-4}, 2\times10^{-2}$], T=2000, a batch size of 40 images, and the Huber loss in place of the L2 loss with the Adam optimizer \citep{kingma2014adam}. We used a learning rate of $5\times10^{-4}$ for the model at 64x64 and $2\times10^{-4}$ for the model at 128x128 and saved checkpoints every 2000 iterations, to enable sampling from multiple models. For the 64x64 case, we train models with 3 different seeds. We use code blocks and architecture from Hugging Face's The Annotated Diffusion model \citep{annotateddiff}, \url{https://github.com/lucidrains/denoising-diffusion-pytorch} and \cite{ho2020denoising}. The architecture we use is similar to that in \cite{ho2020denoising}, and consists of a U-Net \citep{ronneberger2015u} with 4 down and up-sampling blocks consisting of 2 ResNet blocks \citep{zagoruyko2016wide}, group-normalization \citep{wu2018group}, and attention \citep{vaswani2017attention, shen2021efficient}. We use the Weights and Biases framework \citep{wandb} for our experiments. The code for experiments in this paper is available at the following repository: \url{https://github.com/nmudur/diffusion-models-astrophysical-fields-mlps}.

\begin{figure}
\centering
\includegraphics[keepaspectratio=true, width=.46\linewidth]{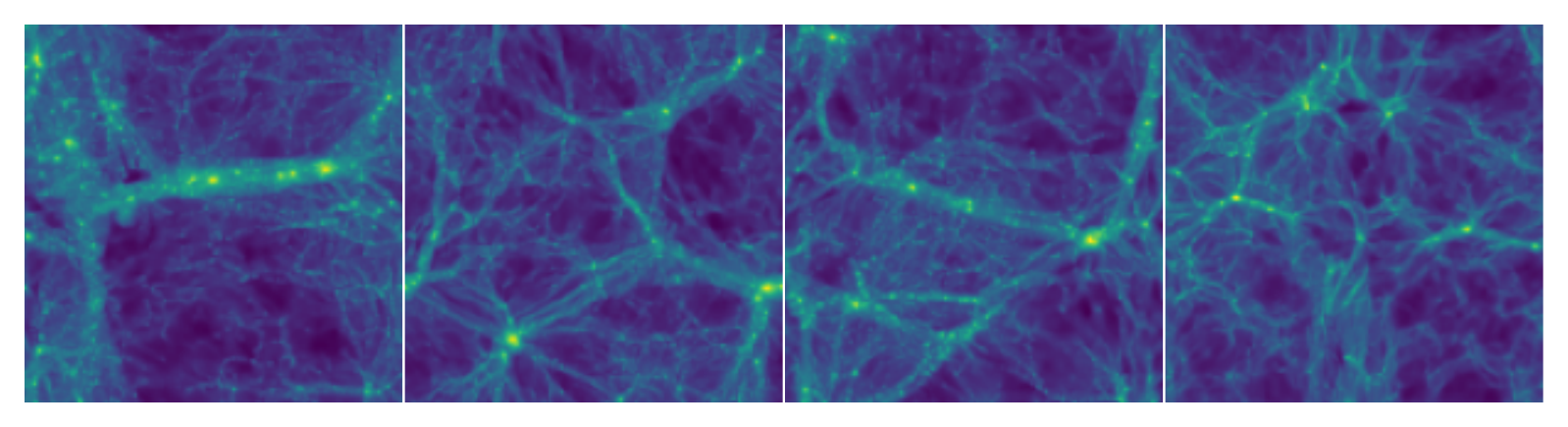}\vrule
\includegraphics[keepaspectratio=true, width=.46\linewidth]{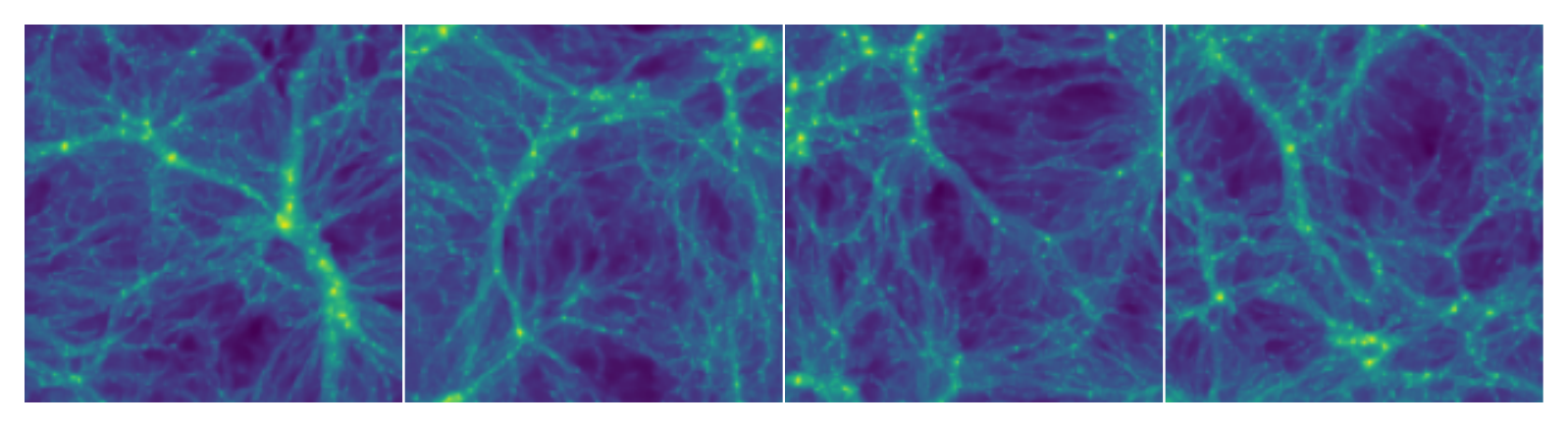}
\includegraphics[keepaspectratio=true, width=.46\linewidth]{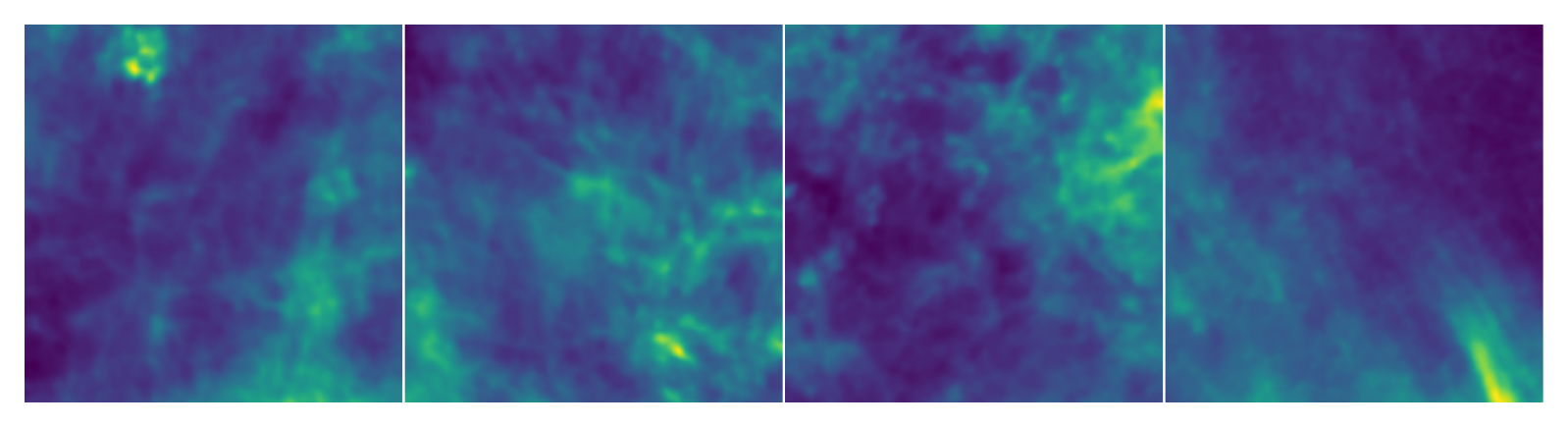}\vrule
\includegraphics[keepaspectratio=true, width=.46\linewidth]{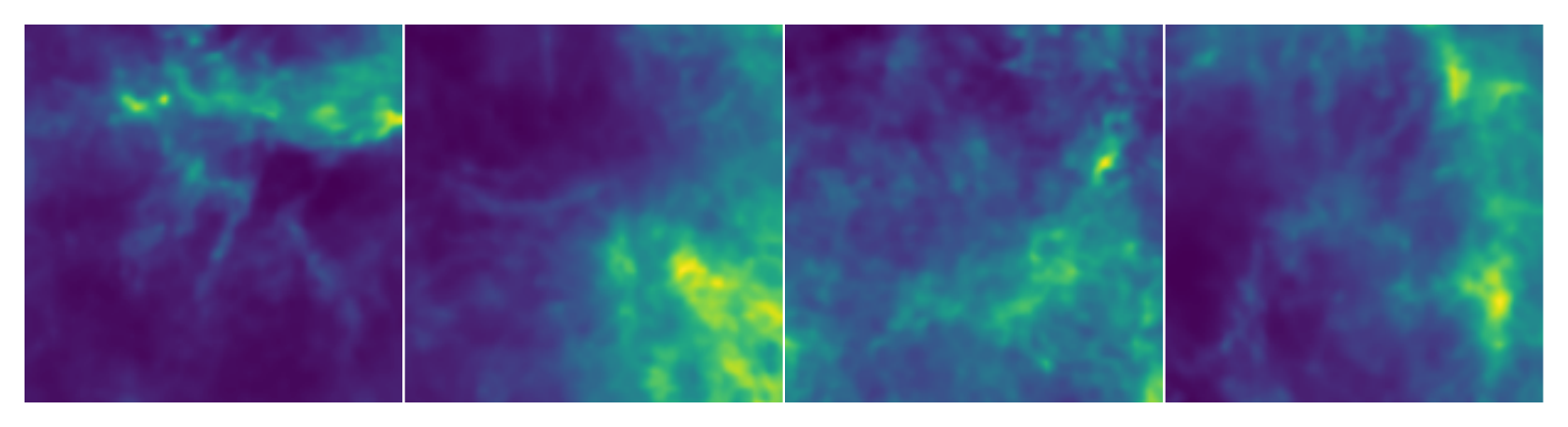}
\caption{Four log cold dark matter mass density fields from the training data (top left) and from the sampled model (top right) at 128x128. Four samples of dust from the training data (bottom left) and from the trained model (bottom right).}
\label{fig:cosmo_samples128}
\end{figure}

\begin{figure}[h]
\centering
\includegraphics[keepaspectratio=true, width=.24\textwidth]{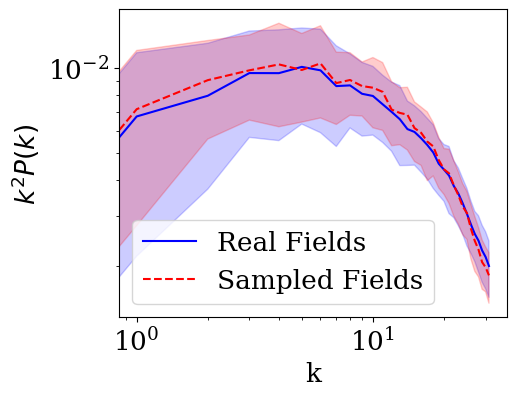}
\includegraphics[keepaspectratio=true, width=.24\textwidth]{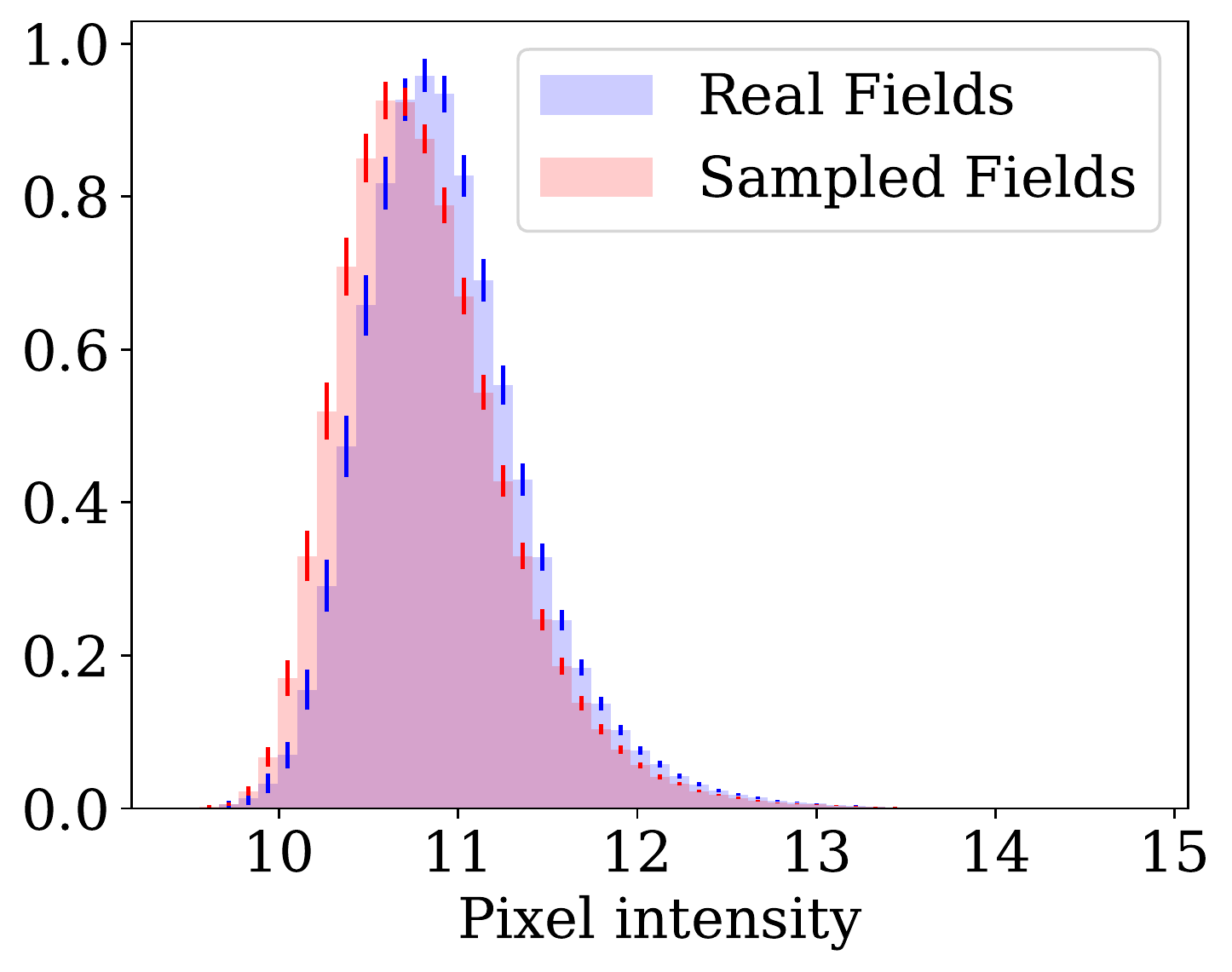}
\includegraphics[keepaspectratio=true, width=.24\textwidth]{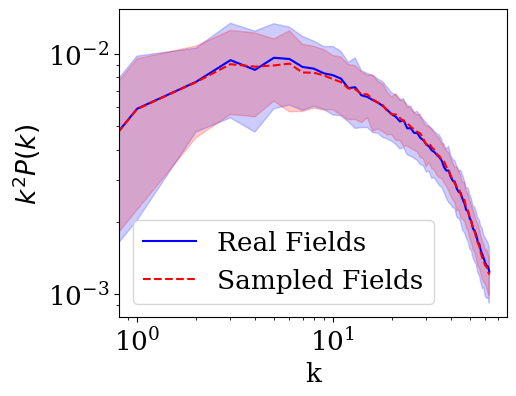}
\includegraphics[keepaspectratio=true, width=.24\textwidth]{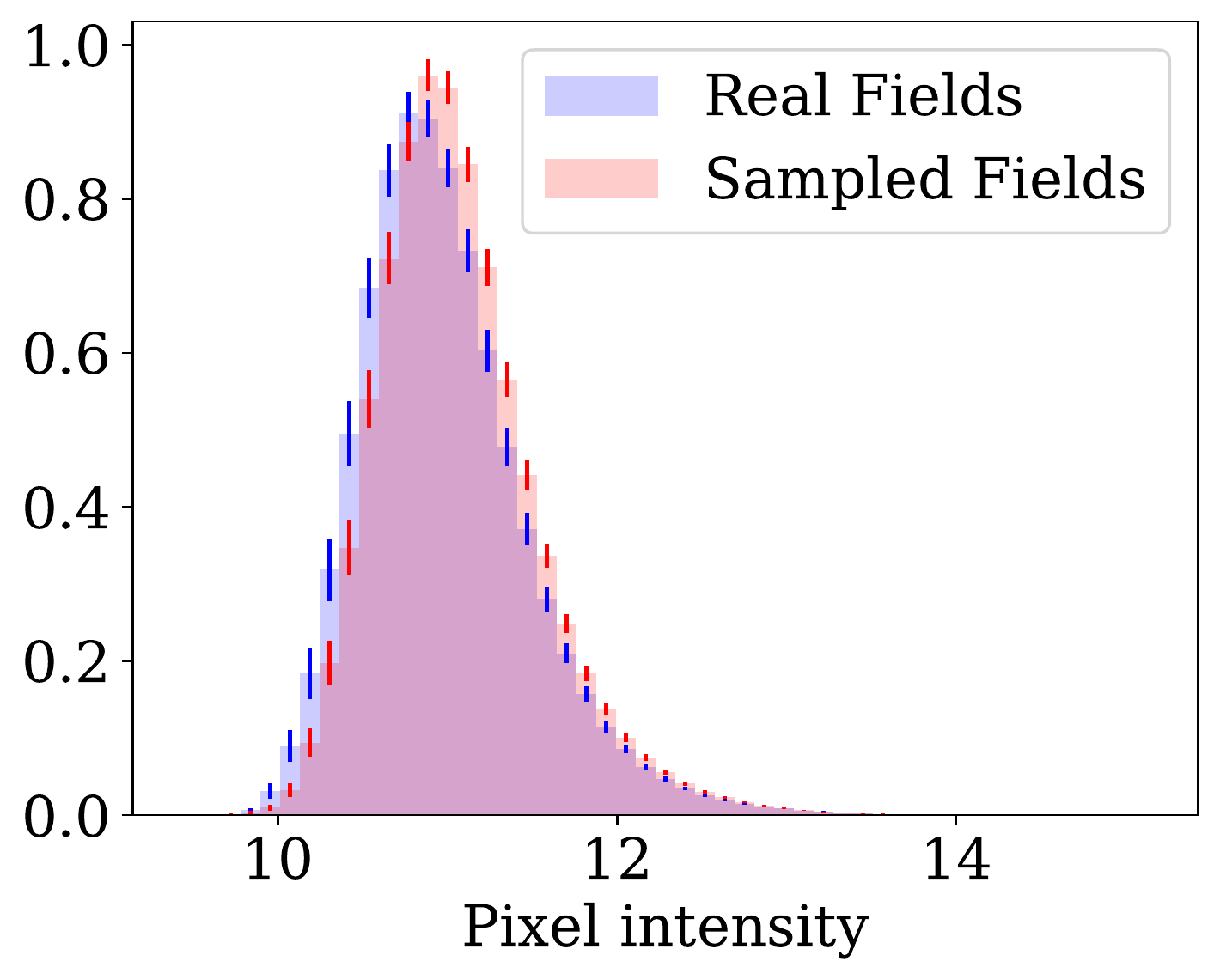} 
\includegraphics[keepaspectratio=true, width=.35\textwidth]{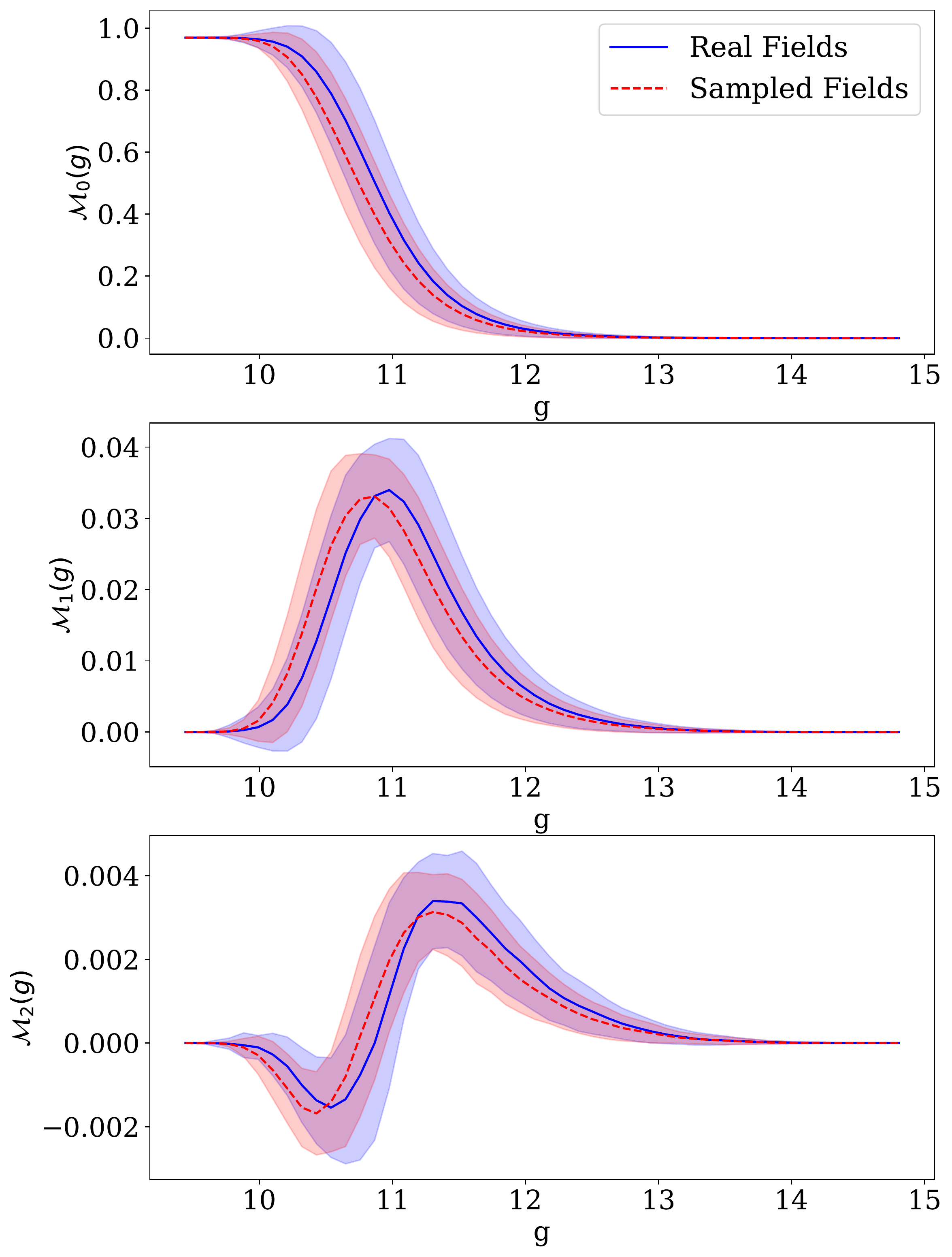} 
\includegraphics[keepaspectratio=true, width=.35\textwidth]{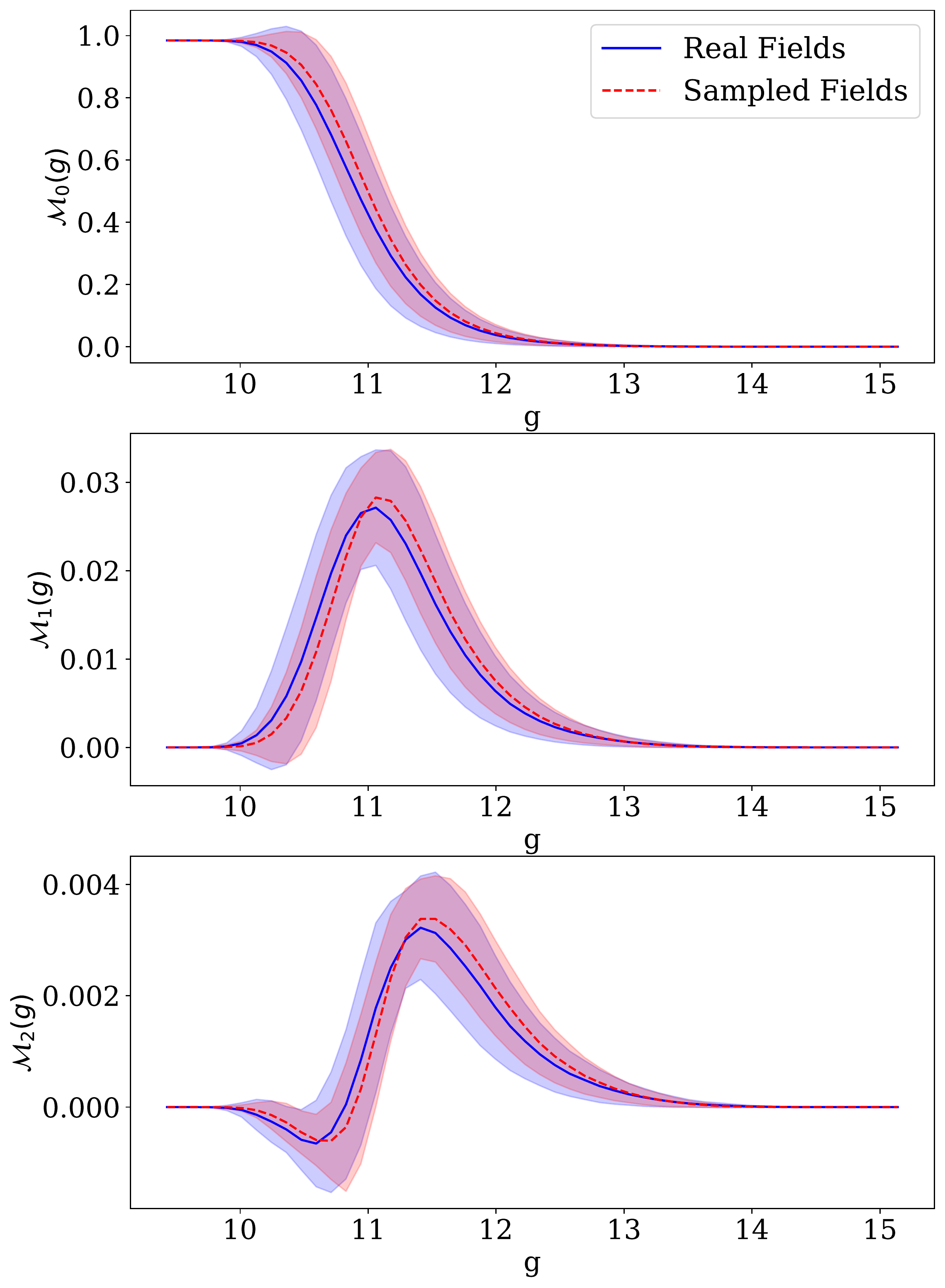} 
\caption{Power spectra, normalized pixel histograms and Minkowski functionals for 100 draws from the real fields and the trained models at 64x64 (left) and 128x128 (right). The envelopes in the power spectra and the Minkowski functionals' panels represent the standard deviation of the value of the statistic in each bin. The height of the bars in the pixel histogram is the mean height of the histogram bin and the error bar is the standard error over all 100 samples.}
\label{fig:cosmo_ss}
\end{figure}

\paragraph{Summary Statistics}
Four samples from the trained model at 128x128 are plotted in the top right panel of Figure \ref{fig:cosmo_samples128}. We consider three summary statistics -- the power spectrum, the normalized pixel intensity histograms and the three Minkowski functionals in Figure \ref{fig:cosmo_ss} to gauge consistency between the sampled fields and the true image distribution. Minkowski functionals \citep{schmalzing1995minkowski,schmalzing1998minkowski} are topological descriptors of fields sensitive to correlation functions beyond the second order and can be used as metrics to gauge how similar the statistics of the generated fields are to those of fields from the true distribution \citep[e.g.,][]{tamosiunas2021investigating, regaldo2022generative}. They are computed as integrals over excursion sets with pixels whose intensity is greater than a value $g$. In 2D, $\mathcal{M}_0, \mathcal{M}_1, \mathcal{M}_2$ reflect the number of pixels in the excursion set (area), the length of its boundary (perimeter) and the number of holes. We use QuantImPy \citep{boelens2021quantimpy} to compute these functionals. 

The loss function used to train these models does not directly enforce the summary statistics of the generated images to exactly match the summary statistics of the training distribution. Thus, while the mean of the statistics of the generated samples across checkpoints typically lie within the one standard deviation envelope of the power spectra and Minkowski functionals of the true distribution, convergence in terms of the loss does not imply that the converged models are stationary with respect to the distribution of the summary statistics of the generated images.  We thus draw 100 samples from each of the last 10 checkpoints of the trained diffusion models and from the real (train) dataset. Since the variability across checkpoints is similar across all three seeds for the 64x64 models, we train a single model for the 128x128 case and sample from its last 10 checkpoints. For the samples across all 10 checkpoints, the mean absolute fractional difference in the power spectrum is typically around 10\% (std. dev. 5\% for 64x64) and 15\% (std. dev. 10\% for 128x128) and is higher at the lowest and highest $k-$bins, which are most affected by cosmic variance and noise, respectively. For both the 128x128 model and all 3 runs of the 64x64 model, at least one checkpoint with a mean (over all bins) absolute fractional difference of less than 5\% could be found. As in \cite{mustafa2019cosmogan} and \cite{tamosiunas2021investigating} that use adherence to summary statistics as a model selection criterion, we identify the checkpoints with the lowest absolute fractional error in the power spectrum and the best agreement with the Minkowski functionals. The statistics plotted in Figure \ref{fig:cosmo_ss} correspond to these models. For the 64x64 case, we plot the statistics for the worst of the three best-case models (one for each seed). While the mean and the spread of the power spectra appear to be largely consistent for these models, the Minkowski functionals are more sensitive to differences between the true and the generated samples. We intend to explore whether these differences and the lack of convergence to a distribution stable with respect to the summary statistics can be mitigated with different design choices, or whether more fundamental changes to the method are required.

\section{A Generative Model for Interstellar Medium Fields}
\begin{figure}
\centering
\includegraphics[keepaspectratio=true, width=0.6\textwidth]{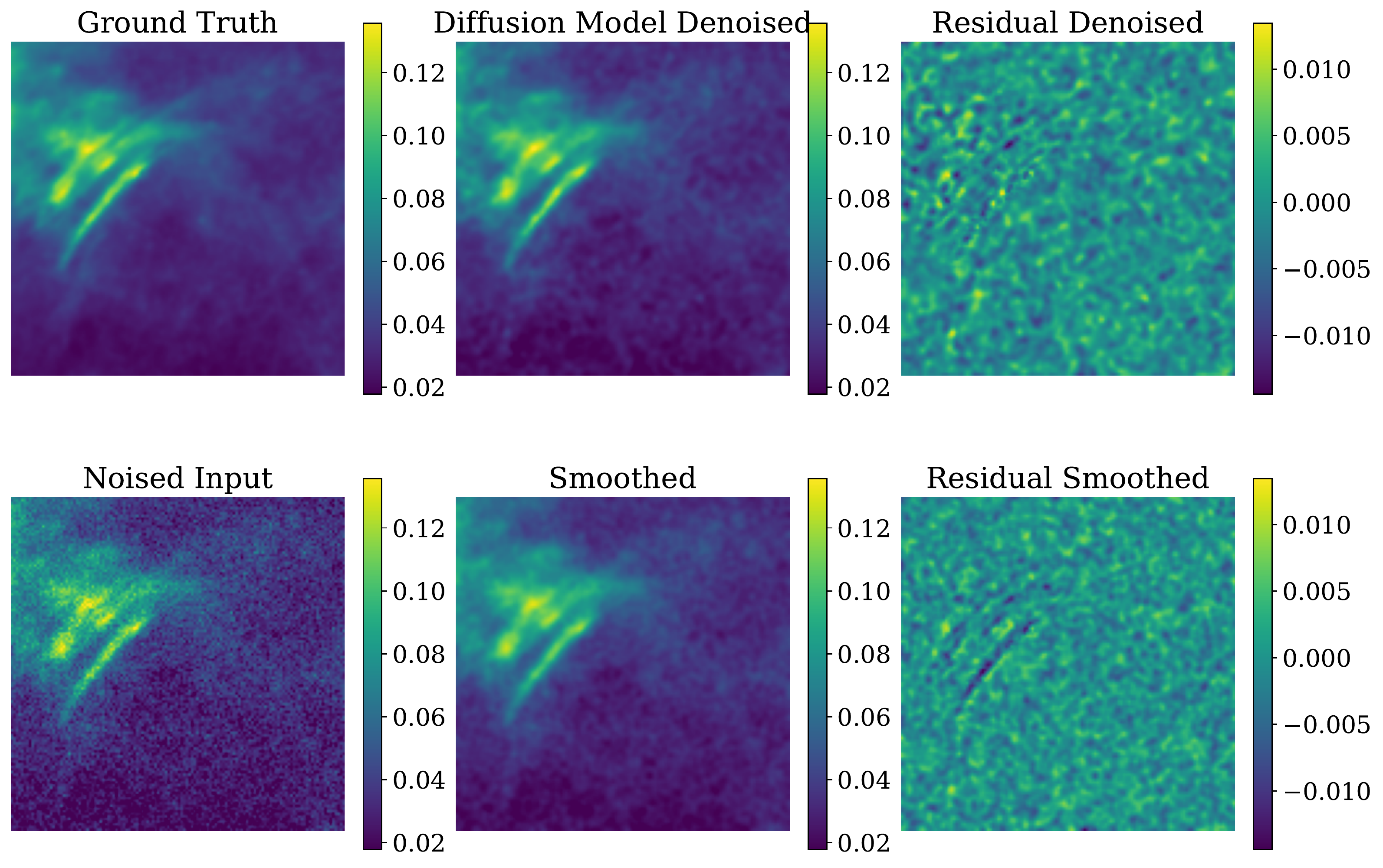}
\caption{Ground truth, noised input, denoised image and its residual (denoised - truth), smoothed baseline and its residual (smoothed - truth).}
\label{fig:denoise}
\end{figure}

\paragraph{Dataset and Training Details} The dataset consists of 12482 images from the Schlegel-Finkbeiner-Davis (SFD) \citep{schlegel1998maps} map of interstellar dust extinction inferred from emission at 100 microns. We restrict ourselves to images with extinction $E_{SFD}<3$. Each image is 128x128 and spans an area of $(6.4^\circ)^2$ on the sky. The validation set consists of images lying in Galactic longitude $0^\circ<l<42^\circ$ and the test set (held out for future validation) consists of images lying between $200^\circ<l<240^\circ$. All other images (roughly 69\%) belong to the train distribution.  The train images are augmented with rotations and flips, yielding 68048 images (8506x8). We apply a minmax transform that scales the minimum and the maximum pixel intensity of \textit{each image} to [-1, 1]. We peg the transform to each image because images of interstellar dust span a larger dynamic range and pixel intensities in two images are much more likely to be dissimilar than for the cosmic web. We train our models for 42k iterations with a learning rate of $6\times10^{-5}$. All other training details are the same as in Section \ref{cosmo}. Four sampled images are plotted in the bottom right panel of Figure \ref{fig:cosmo_samples128}. For both the generated cosmic web images and the dust images we see that the models are able to capture a rich variety of structure. 
\paragraph{Denoising Tests}
We select a filamentary field from the validation dataset, whose standard deviation corresponds to roughly the $50^{th}$ percentile of the standard deviation across all images in the validation set. We add $\mathcal{N}(0, \sigma^2)$ noise such that $\sigma$ is 20\% of the mean of the intensity in the image. The noisy input is scaled to [-1, 1] and the timestep at which $\sqrt{1-\bar{\alpha_t}}$ is closest to the corresponding scaled sigma $\sigma_{tr}$ is identified for each image. We then iteratively sample from $p_\theta(x_{t-1}|x_t)$ from $t=t_{\sigma_{tr}}$ to $t=0$ to derive the denoised image. As a baseline, we find the corresponding Gaussian filter that would reduce the RMSE in the low-signal portion on the bottom right of the image by the same factor (3.3). Figure \ref{fig:denoise} plots the denoised images with the model and the baseline. The correlation of the residual with filaments is significantly lower with the diffusion model than with the baseline. 

\section{Conclusions and Future Work}
In this work, we investigated whether DDPMs are able to learn and sample from the image distribution for two astrophysical fields — one from simulations and the other from real interstellar dust maps — in a ‘physically meaningful’ way, gauged by two different yardsticks.  In the case of models trained to generate images of interstellar dust, the models are able to denoise highly filamentary images of interstellar dust while recovering underlying structure better than a smoothing baseline. Approaches such as \cite{regaldo2022generative} involve learning generative models of interstellar dust using the wavelet statistics of a single image, whereas our description learns a prior using multiple images. The former approach can be useful in cases where limited training data is available while the latter approach is more likely to account for diversity and multiple modes of the image distribution. In the context of dark matter density fields from simulations, we examined three sets of summary statistics of cosmological significance. The ability to generate images with the same statistics is an important first step toward deploying these models as emulators. While DDPMs show promise in terms of our ability to find models that generate samples that are consistent up to 10\% in the power spectrum, we intend to work on finding architectures or models with inductive biases that prioritize convergence to distributions that are stationary with respect to these summary statistics.  

\section{Broader Impact}
Several papers have examined and improved the performance of score-based generative models for standard machine learning datasets using standardized metrics such as the Fr\'echet Inception Distance \citep{heusel2017gans}, demonstrated their ability to generate photo-realistic images conditional on text inputs \citep{ramesh2022hierarchical} and identified issues to work on, such as their slower sampling speed, relative to GANs. This raises the question -- how can these models help accelerate science? As a first demonstration of the use of these models to generate fields from the interstellar medium, we hope that this paper can motivate both potential use cases for these models in astrophysics and cosmology as well as work on imbuing these models with physical inductive biases that make them more suitable for applications where practitioners care about recovering specific summary statistics. While our focus here has been on astrophysics, the questions we consider and the desire for high-fidelity generative models capable of sampling distributions of images faster than simulations have wider applications across the physical sciences  \citep{kasim2021building} — from the geophysical sciences to high energy physics \citep{paganini2018accelerating}.

\section{Acknowledgements}
We thank Shuchin Aeron, Carolina Cuesta-Lazaro, Tanveer Karim, Andrew K. Saydjari, and Justina R. Yang for helpful discussions. This work was supported by the National Science Foundation under Cooperative Agreement PHY2019786 (The NSF AI Institute for Artificial Intelligence and Fundamental Interactions). 

\bibliographystyle{unsrtnat}
\bibliography{sample}



\appendix

\section{Appendix}
\subsection{Experiments and Compute Time}
The 64x64 model took 1.25h to train for 60k iterations while the 128x128 models took 13.5 hours to train for the same number of iterations. We used an NVIDIA A100 for most experiments and runs.
We ran most experiments on the 64x64 model, since it took 1.25 hours to train for 60k iterations. The cumulative compute time spent on experiments was around 5 days. Sampling 10 images from the trained diffusion models took $\sim40$ seconds for the 64x64 model and $\sim110$ seconds for the 128x128 model on the A100. We gauged whether other hyperparameter choices were better by plotting the summary statistics of samples from the generated model (as in Figure \ref{fig:cosmo_ss}).
\begin{itemize}
\item We tried the cosine learning schedule proposed in \citep{nichol2021improved} and found the linear schedule to work better, for the number of iterations we experimented with.
\item We chose a lower learning rate for the dust images since we found this to be more stable.
\item We also tried learning rates of  [$6\times10^{-5}$, $2\times10^{-4}$, $1\times10^{-3}$] for the 64x64 model and did not find any of the other learning rates to improve performance.
\item We examined the summary statistics for samples from the 64x64 runs described in Section 3 for the last 15 checkpoints and did not find significant differences in the quality of the summary statistics.
\end{itemize}


\end{document}